\documentclass[aps,prl,twocolumn,longbibliography,10pt]{revtex4-2}

\usepackage{amsmath}
\usepackage{amssymb}
\usepackage{graphicx}
\usepackage[colorlinks=true,citecolor=blue,urlcolor=cyan]{hyperref}
\usepackage{bm}
\usepackage{subfigure}
\usepackage{xcolor}
\usepackage{enumitem}
\usepackage{amsthm}

\newcommand{\Z}{\mathbb{Z}}

\newcommand{\EE}{\mathcal{E}}
\newcommand{\NN}{\mathcal{N}}
\newcommand{\HH}{\mathcal{H}}
\newcommand{\OO}{\mathcal{O}}
\newcommand{\Tr}{\mathrm{Tr}}
\newcommand{\sgn}{\mathrm{sgn}}

\definecolor{darkblue}{rgb}{0.0,0.28,0.85}

\makeatletter
\newenvironment{theorem*}{%
  \begin{trivlist}%
    \item[\hskip\labelsep\bfseries Theorem\@addpunct{.}]\itshape
}{%
  \end{trivlist}%
}
\makeatother

\makeatletter
\newenvironment{lemma*}{%
  \begin{trivlist}%
    \item[\hskip\labelsep\bfseries Lemma\@addpunct{.}]\itshape
}{%
  \end{trivlist}%
}
\makeatother

\begin{document}

\title{ Tensor-Network Algorithm for Many-Body Trace Norms}

\author{Seunghun Lee}
\affiliation{Department of Physics, Korea Advanced Institute of Science and Technology, Daejeon 34141, Republic of Korea}
\author{Eun-Gook Moon}
\thanks{egmoon@kaist.ac.kr}
\affiliation{Department of Physics, Korea Advanced Institute of Science and Technology, Daejeon 34141, Republic of Korea}

\begin{abstract}
    Trace norms are fundamental to quantum information theory, yet in many-body systems their evaluation remains a major computational bottleneck, as it generally requires diagonalizing exponentially large operators. Here, we overcome this bottleneck by introducing a controlled tensor-network algorithm for estimating the trace norm of matrix product operators without full diagonalization. The key idea is to combine Zolotarev's rational approximation to the sign function with a variational formulation solved using a density-matrix-renormalization-group-like algorithm. The resulting approximation is systematically improvable, with its accuracy controlled by the rational approximation parameters and the spectral weight near zero. Beyond the reach of exact diagonalization, we demonstrate controlled trace-norm calculations for entanglement negativity, quantum fidelity and quantum Fisher information, achieving substantially improved accuracy over polynomial-based Lanczos approaches. Our results establish trace-norm-based quantities as practical tensor-network observables, opening a route toward tensor-network studies of quantum information in mixed states.
\end{abstract}

\maketitle

\textit{Introduction}---Tensor-network methods are indispensable for studying quantum many-body systems, providing compact and physically motivated representations of states and operators whose full Hilbert-space descriptions grow exponentially with system size~\cite{schollwock2011density,cirac2021matrix,orus2014a}. In one dimension, a prominent tensor-network ansatz for pure states is the matrix product state (MPS)~\cite{fannes1992finitely,vidal2003efficient,perez2007matrix}, which efficiently captures the entanglement structure underlying the success of numerical algorithms such as the density-matrix renormalization group (DMRG)~\cite{white1992density,white1993density,mcculloch2007density,crosswhite2008finite}. 

Beyond pure states, various physically relevant operators---including thermal states, decohered mixed states, reduced density matrices, and many-body operators such as Hamiltonians---can be efficiently represented using matrix product operators (MPOs)~\cite{verstraete2004matrix,pirvu2010matrix,zwolak2004mixed}. MPOs provide a powerful framework for manipulating operators in many-body systems, enabling the simulation of thermal states and open-systems. Their usefulness is further supported by theoretical results showing that one-dimensional mixed states with sufficiently low entanglement admit efficient MPO approximations~\cite{jarkovsky2020efficient}. Moreover, a recent protocol for learning mixed states directly in the MPO representation~\cite{votto2026learning} suggests that MPO descriptions are  becoming experimentally accessible, motivating the development of numerical methods for extracting nontrivial physical quantities directly from MPOs.

In this work, we focus on a fundamental quantity in quantum information theory: the trace norm $\| A \|_1$ of an MPO $A$. The trace norm underlies several key information-theoretic measures, including the trace distance $d(\rho, \sigma) = \| \rho - \sigma \|_1 / 2$, which quantifies state distinguishability~\cite{nielsen2001quantum}, and the entanglement negativity $\NN(\rho) = (\| \rho^{T_R} \|_1 - 1) / 2$, a widely used computable measure of mixed-state entanglement~\cite{vidal2002computable,plenio2005logarithmic}. However, because the trace norm involves a non-analytic function (the absolute value) of the spectrum, its evaluation in many-body systems typically requires full diagonalization of an exponentially large matrix. Moreover, there is no polynomial-time algorithm that exactly computes the trace norm of a general MPO (assuming $\mathrm{P} \neq \mathrm{NP}$)~\cite{kliesch2014matrix,note_one}. Nevertheless, this does not preclude practical estimation for physically relevant MPOs, motivating the search for controlled and tractable approximation schemes.

\begin{figure}[t]
    \includegraphics[width=\columnwidth]{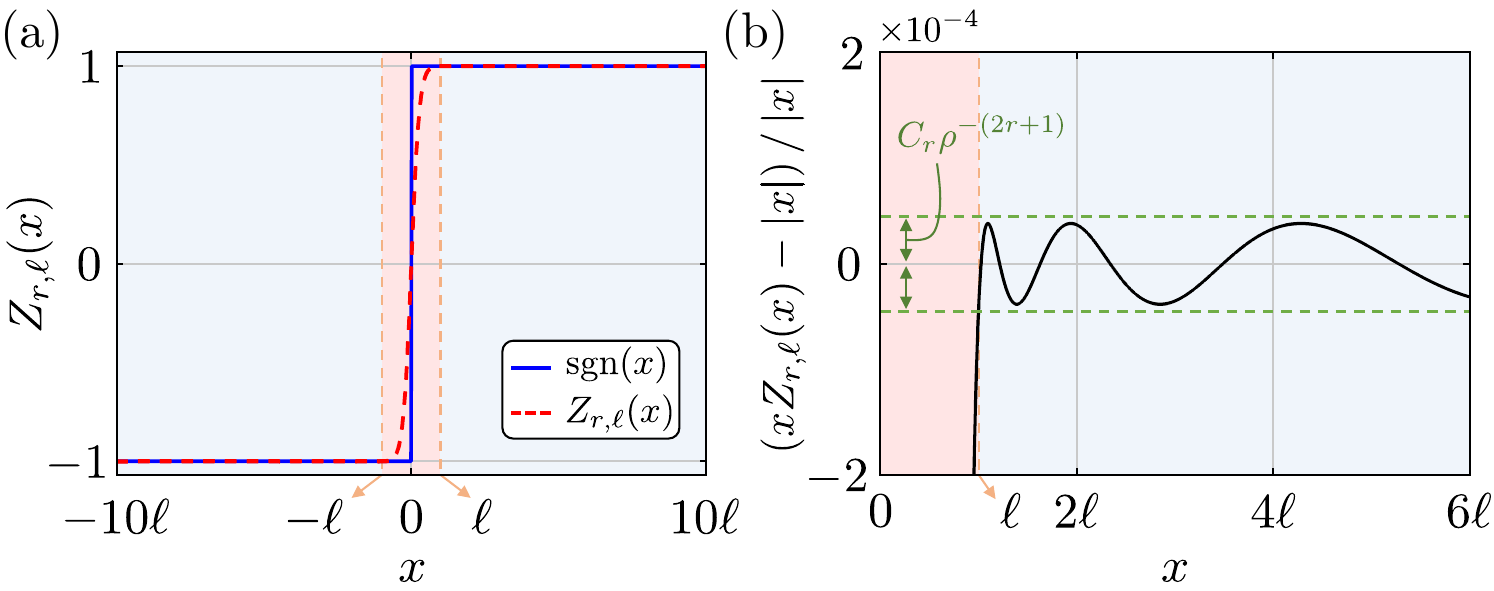}
    \caption{(a) The sign function $\sgn(x)$ (blue solid curve) and Zolotarev's rational approximation $Z_{\ell,r}(x)$ (red dashed curve). The approximation works well on $[1, -\ell] \cup [\ell, 1]$ (blue-shaded region) and breaks down for $|x| < \ell$ (red-shaded region). (b) Relative error of the absolute-value approximation $x Z_{\ell,r}(x)$, bounded by a constant $C_r \rho^{-(2r+1)}$ (green dashed line) within the blue-shaded region. The parameters used are $\ell = 10^{-7}$ and $r = 20$.}
    \label{fig:Zolotarev}
\end{figure}

Here, we introduce a tensor-network method for estimating the trace norm of moderate-sized MPOs without full diagonalization, applicable also to non-Hermitian MPOs. Our approach builds on Zolotarev’s rational approximation to the sign function, which offers exponentially accurate approximations with two tunable parameters. Applying this to MPOs, we reformulate trace-norm estimation as a variational optimization problem that can be solved using a DMRG-like algorithm. This yields a controlled approximation to various trace-norm-based quantities, with an explicit error bound governed by the approximation parameters and the low-lying spectral weight. In contrast to polynomial-based approaches such as Lanczos methods, our framework provides improved accuracy and controllability for the trace norm.

\begin{figure*}[t]
    \centering
    \includegraphics[width=2\columnwidth]{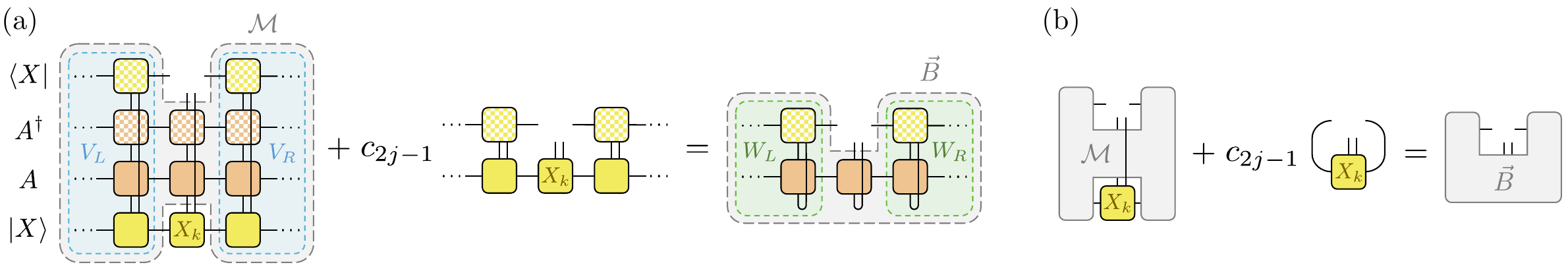}
    \caption{(a) Local linear system obtained by differentiating the objective function Eq.~\eqref{reformulation} with respect to complex-conjugate of the local tensor $X_k$ of the variational MPS $| X \rangle$. The orange and yellow boxes depict the MPO $A$ and the variational MPS $| X \rangle$, respectively, while the checkered patterns indicate Hermitian conjugation. The matrix $\mathcal{M}$ and vector $\vec{B}$ are constructed from the left and right environment tensors $V_{L,R}$ and $W_{L,R}$, analogous to DMRG. (b) The simplified local linear system [see Eq.~\eqref{locLE}] obtained after bringing $| X \rangle$ into the mixed canonical form.}
    \label{fig:DMRGlike}
\end{figure*}

We benchmark our method by estimating the entanglement negativity of decohered cluster states and random separable mixed states. We further demonstrate that the same framework enables the computation of quantum fidelity between mixed states---an important yet generally intractable quantity---when their purifications are available as MPSs. We illustrate this extension by estimating the quantum Fisher information of a noisy GHZ state from fidelity estimates. Overall, our approach provides a practical route to estimating trace-norm–based quantities for MPOs, beyond the regime accessible to exact diagonalization.

\textit{Zolotarev's Rational Approximation}---The trace norm of a matrix $A$ is defined by $\| A \|_1 := \Tr [|A|] = \sum_i \sigma_i$, where $\sigma_i$ are the singular values of $A$. For a Hermitian $A$, $\| A \|_1$ equals the sum of the absolute eigenvalues of $A$. Since the absolute-value function can be written as $|x| = x \cdot \sgn(x)$, it is natural to consider Zolotarev's rational approximation $Z_{\ell,r} (x)$ for the sign function $\sgn(x)$ on $[-1, -\ell] \cup [\ell, 1]$~\cite{zolotarev1877application,achieser2013theory,nakatsukasa2016computing}:
\begin{align} \label{Zolotarev}
    Z_{\ell,r} (x) := M x \left( 1 + \sum_{j=1}^r \frac{a_j}{x^2 + c_{2j-1}} \right),
\end{align}
where $M$, $a_j$ and $c_{2j-1}$ are certain positive constants. The approximation Eq.~\eqref{Zolotarev} is characterized by two parameters, $\ell \in (0, 1)$ and $r \in \Z_{\geq 0}$. The parameter $\ell$ sets the spectral cutoff below which the approximation is not applied, and $r$ controls the degree of approximation; the approximation error decays exponentially with $r$~\cite{nakatsukasa2016computing}:
\begin{align} \label{ZolotarevErr}
    \max_{x\in [-1,-\ell] \cup [\ell, 1]} | \sgn(x) - Z_{\ell,r} (x) | = C_{r} \rho^{-(2r+1)},
\end{align}
where $\rho > 1$ and $1 \leq C_r \leq \frac{2}{1 - \rho^{-1}}$ are some constants. We give explicit formulae for the constants in the End Matter to be self-contained. Thus, taking sufficiently small $\ell$ and large $r$ makes $Z_{\ell,r}(x)$ a good approximation for the sign function as illustrated in Fig.~\ref{fig:Zolotarev}. In the Supplemental Material (SM), we discuss the exponential advantage of rational over polynomial approximations for $\sgn(x)$, with the latter underlying Lanczos-based methods, measured by the number of terms required to achieve a given accuracy.

Now, we consider a Hermitian MPO $A$ with its spectral radius bounded above by unity (i.e., $|A| \preceq 1$), which can be achieved by normalizing $A$ by its extremal eigenvalues that are obtainable via DMRG. From Eq.~\eqref{Zolotarev}, we propose the following estimator for the trace norm of an MPO $A$:
\begin{equation} \label{MyEstimator}
    \begin{aligned} 
        f_{\ell,r}(A) &:= \Tr[A \cdot Z_{\ell,r}(A)] \\
        &= M \left( \Tr[A^2] + \sum_{j=1}^r a_j \Tr \left[ \frac{A^2}{A^2 + c_{2j-1} I} \right] \right).
    \end{aligned}
\end{equation}
From Eq.~\eqref{ZolotarevErr}, we obtain the relative error bound
\begin{align} \label{MyEstimatorErr}
    \frac{\left| f_{\ell,r} (A) - \| A \|_1 \right|}{\| A \|_1} &\leq C_r \rho^{-(2r+1)} + R_{<\ell},
\end{align}
where $R_{<\ell} := \| A \|_1^{-1} \sum_{i: |\lambda_i| < \ell} |\lambda_i|$ is the fraction of the trace norm contributed by absolute eigenvalues of $A$ less than $\ell$, thus reflecting the truncated spectral weight near zero (see the End Matter for the derivation). Note that potential additional errors that might appear during the tensor-network computation are ignored in Eq.~\eqref{MyEstimatorErr}, and will be discussed later.

Our estimator Eq.~\eqref{MyEstimator} can be straightforwardly extended to non-Hermitain MPOs via a simple modification: when an $N$-qubit MPO $A$ is non-Hermitian, consider the following $(N+1)$-qubit Hermitian matrix:
\begin{align} \label{NonHerm2Herm}
    A' = \begin{pmatrix}
        0 & A \\
        A^\dagger & 0
    \end{pmatrix} = | 0 \rangle \langle 1 | \otimes A + | 1 \rangle \langle 0 | \otimes A^\dagger.
\end{align}
If $A$ is an MPO with bond dimension $D$, then $A'$ admits an MPO representation with bond dimension at most $2D$. One can easily show that $\| A \|_1 = \frac 12 \| A' \|_1$. Therefore, we can apply Eq.~\eqref{MyEstimator} to the Hermitian MPO $A'$ (with proper normalization) to compute the trace norm of the non-Hermitian MPO $A$. 

\textit{Reformulation as a DMRG-Like Problem}---A remaining challenge in Eq.~\eqref{MyEstimator} is to evaluate the trace terms $\Tr [A^2 (A^2 + c_{2j-1} I)^{-1}]$. (The term $\text{Tr}[A^2]$ is computed straightforwardly.) A key idea is to rewrite these trace terms as matrix elements $\langle A | [I \otimes (A^2 + c_{2j-1} I)]^{-1} | A \rangle$, where $| A \rangle := \mathrm{vec}(A)$ denotes the vectorization of $A$. Since $A^2 + c_{2j-1} I$ is Hermitian and positive definite, this matrix element can be expressed as the following optimization problem:
\begin{align} \label{reformulation}
    \hspace{-4pt} \max_{| X \rangle} \left( \langle A | X \rangle + \langle X | A \rangle - \langle X | [I \otimes (A^2 + c_{2j-1} I)] | X \rangle \right),
\end{align}
where the maximization is over all states $|X\rangle$ in the doubled Hilbert space.

We variationally solve the optimization problem in Eq.~\eqref{reformulation} by restricting the search space for $|X\rangle$ to the space of MPSs with bond dimension $\chi$. Taking the partial derivative of the objective function in Eq.~\eqref{reformulation} with respect to local tensor $X_k$ of the variational MPS $| X \rangle$, we obtain the local linear system shown in Fig.~\ref{fig:DMRGlike}(a) for each $1 \leq j \leq r$. The matrix $\mathcal{M}$ and vector $\vec{B}$ in Fig.~\ref{fig:DMRGlike}(a) can be constructed by storing and updating the environment tensors $V_{L,R}$ and $W_{L,R}$ as in the standard DMRG algorithm. Using the mixed canonical form of the MPS $| X \rangle$~\cite{schollwock2011density}, the local linear system simplifies to 
\begin{align} \label{locLE}
    (I \otimes \mathcal{M} + c_{2j-1} I \otimes I) \vec{X}_k = \vec{B},
\end{align}
where $\vec{X}_k$ is the vectorized local tensor $X_k$ [see Fig.~\ref{fig:DMRGlike}(b)]. We then apply a DMRG-like sweeping algorithm that sequentially solves these local linear equations and updates the MPS $|X \rangle$ until relative change in the estimated trace value falls below a prescribed threshold $\epsilon_{\mathrm{th}}$. (We used $\epsilon_{\mathrm{th}} = 10^{-6}$ for all benchmarks below.) One can parallelize over $j$ to reduce the computation time. Since the minimal required bond dimension $\chi$ is hard to determine \emph{a priori}, we increase $\chi$ sequentially, seeding each step with the converged MPS $| X\rangle$ from the previous $\chi$, and verify convergence of the estimated trace. For the benchmarks presented below, we verified such convergence, with representative results provided in the SM~\cite{SM}.

We consider two approaches to solving the local linear equation Eq.~\eqref{locLE}. The first is the conjugate gradient method, which is matrix-free, i.e., it does not explicitly construct the $d\chi^2 \times d\chi^2$ matrix $\mathcal{M}$. Its time complexity is $\mathcal{O}(N_{\mathrm{itr}} \chi^3)$, where $N_{\mathrm{itr}}$ is the number of iterations. However, since many-body operators often have ill-conditioned spectra, $\mathcal{M}$ can exhibit a large condition number $\kappa$, requiring $N_{\mathrm{itr}} \sim \sqrt{\kappa}$ iterations for the conjugate gradient method to converge. Alternatively, we find that explicitly constructing $\mathcal{M}$ and solving Eq.~\eqref{locLE} directly (e.g., via Cholesky factorization) is faster and more accurate in practice, with time complexity $\mathcal{O}(\chi^6)$ in solving the linear system.

Despite practical efficiency, our method can face limitations for very large systems due to the possible exponential suppression of small absolute eigenvalues. As the system size grows, it is, in principle, desirable to decrease $\ell$ to resolve the relevant spectral weight. However, $\ell$ is ultimately constrained by machine precision, which can make part of the spectrum difficult to access. Nevertheless, the method remains effective for moderate-sized MPOs and allows exploration of mixed-state regimes well beyond the reach of exact diagonalization, as demonstrated in the benchmarks below. 

\textit{Benchmark: Decohered Cluster State}---To benchmark our method, we estimate the entanglement negativity of a decohered quantum spin chain, which has an exact analytic result for a stringent numerical test. Concretely, we consider a cluster state $| \psi \rangle$ on a one-dimensional open chain, which is the ground state of the Hamiltonian $H = -\sum_{i=2}^{L - 1} Z_{i-1} X_i Z_{i+1} - X_1 Z_2 - Z_{L-1} X_L$~\cite{briegel2001persistent}. Under the phase-flip noise channel $\EE_Z = \bigotimes_i \EE_{Z, i}$ with $\EE_{Z, i}[\cdot] = (1-p) (\cdot) + p Z_i (\cdot) Z_i$ ($p$: error rate), the pure state $| \psi \rangle$ decoheres into a mixed state $\rho = \EE_Z [| \psi \rangle \langle \psi |]$. We show in the SM that the entanglement negativity between the two halves of the chain (with $R$ denoting one half) is given by
\begin{align} \label{Neg_DecoCluster}
    \NN(p) = \left( p^2 - 2p + \frac 12 \right) \Theta (p_c - p),
\end{align}
independent of the chain length $L$~\cite{SM}. Here $\Theta(\cdot)$ denotes the Heaviside theta function and $p_c = 1 - 1 / \sqrt{2}$. 

The decohered mixed state $\rho$ has an exact MPO representation with bond dimension 4. The partial-transposed density matrix $\rho^{T_R}$ then admits the same MPO representation, with the physical legs corresponding to bras and kets swapped in subsystem $R$. Unlike existing approaches whose efficiency relies on a specific bipartition geometry~\cite{hauru2018uhlmann}, our method allows arbitrary bipartitions by simply swapping the physical legs in the relevant subsystem, without rebuilding the tensor network.

\begin{figure}[t]
    \includegraphics[width=\columnwidth]{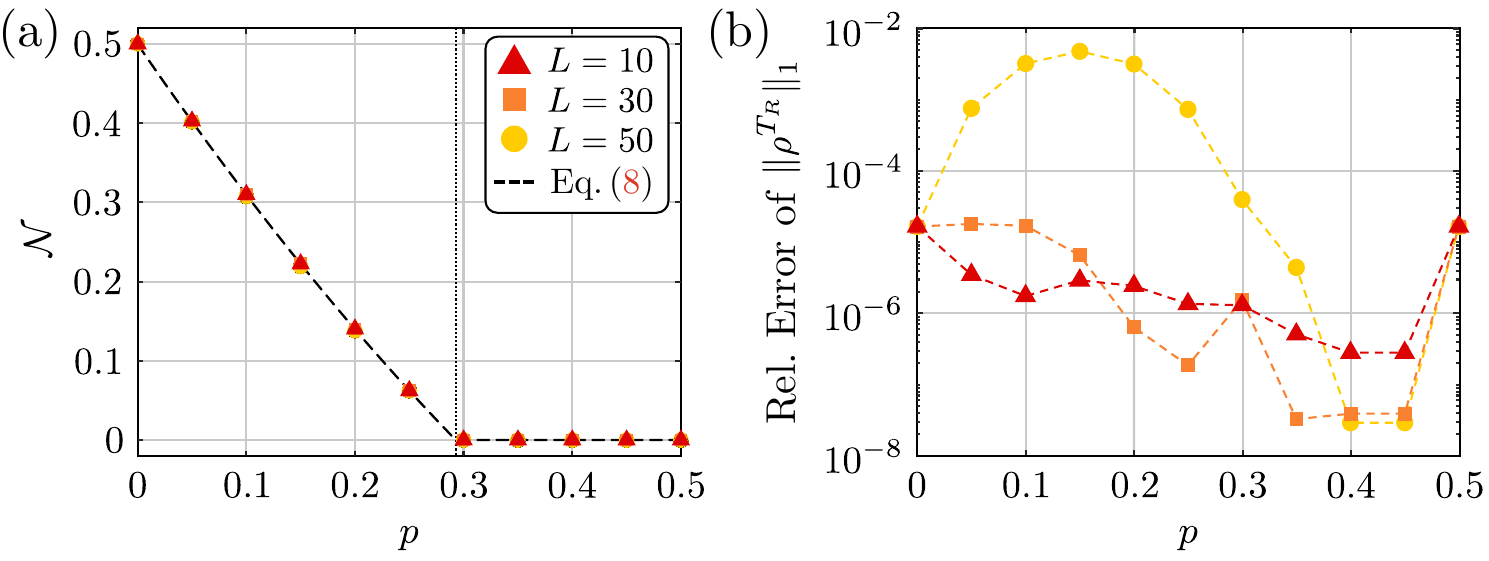}
    \caption{(a) Estimated half-chain entanglement negativity $\NN$ of the decohered cluster state $\rho$ of length $L$ under the phase-flip noise. The dashed curve shows the exact value in Eq.~\eqref{Neg_DecoCluster}. The dotted vertical line marks $p_c = 1 - 1/\sqrt{2}$. (b) Relative error of the estimated trace norm $\| \rho^{T_R} \|_1$. The parameters used are $\ell = 10^{-10}$, $r = 30$ and $\chi = 50$.}
    \label{fig:DecoCluster_Neg}
\end{figure}

Figure~\ref{fig:DecoCluster_Neg}(a) displays the estimated entanglement negativity for various chain lengths $L$. The estimates agree well with the exact analytic result in Eq.~\eqref{Neg_DecoCluster} for all considered $L$, and our method successfully captures the finite-to-zero transition of negativity across $p = p_c$. Figure~\ref{fig:DecoCluster_Neg}(b) shows the relative error in $\| \rho^{T_R} \|_1$, which remains reasonably small across all $p$. We observe that for fixed $\ell$ and $r$ the relative error increases with $L$, reflecting the growth in the number of absolute eigenvalues of $\rho^{T_R}$ below $\ell$ with system size $L$, a generic feature of many-body systems. In the SM, we demonstrate that the relative error decreases systematically and ultimately saturates as the bond dimension $\chi$ increases, confirming the controlled convergence of our approach~\cite{SM}. We also compare our method with the MPO Lanczos method of Refs.~\cite{august2017on,august2018efficient} in the SM, demonstrating a clear advantage in the accuracy and controllability~\cite{SM}.

\textit{Benchmark: Random Separable Mixed States}---As a second benchmark, we test our method on a more general class of MPOs where exact diagonalization is intractable, estimating the entanglement negativity of mixed states $\rho = \sum_{i=1}^q p_i \sigma_{i,1} \otimes \sigma_{i,2}$, where $\sigma_{i,1/2}$ are random density matrices and $\sum_{i=1}^q p_i = 1$. By construction, $\rho$ is separable and thus has zero entanglement. We consider instances with chain length $L = 20$ and $q$ randomly chosen from $1$ to $3$, resulting in MPO bond dimensions up to $D \leq 12$. Notably, for $L = 20$, exact diagonalization of the density matrix is already computationally prohibitive. 

\begin{figure}[t]
    \includegraphics[width=0.8\columnwidth]{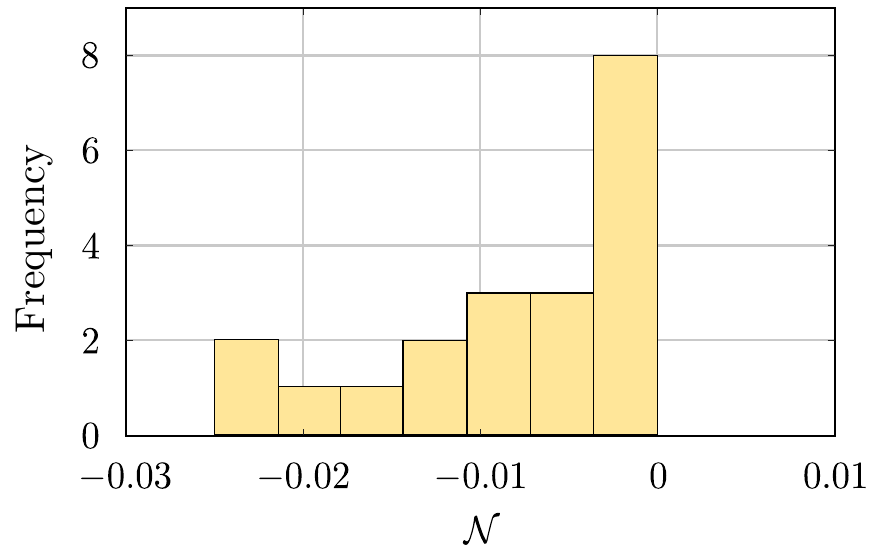}
    \caption{Histogram of the estimated negativity for 20 instances of random separable mixed states $\rho = \sum_{i=1}^q p_i \sigma_{i,1} \otimes \sigma_{i,2}$ with chain length $L = 20$ and $q\leq 3$. The parameters used are $\ell = 10^{-7}$, $r = 20$ and $\chi = 50$.}
    \label{fig:RandomSep_Neg}
\end{figure}

Applying our method to these random separable density matrices, we obtain the histogram shown in Fig.~\ref{fig:RandomSep_Neg}. The estimated negativities are consistently close to zero, in agreement with separability. Small negative deviations arise from the finite $\ell$ and limited bond dimension, and can be systematically reduced by choosing better parameters with more computational time.

\begin{figure}[t]
    \includegraphics[width=\columnwidth]{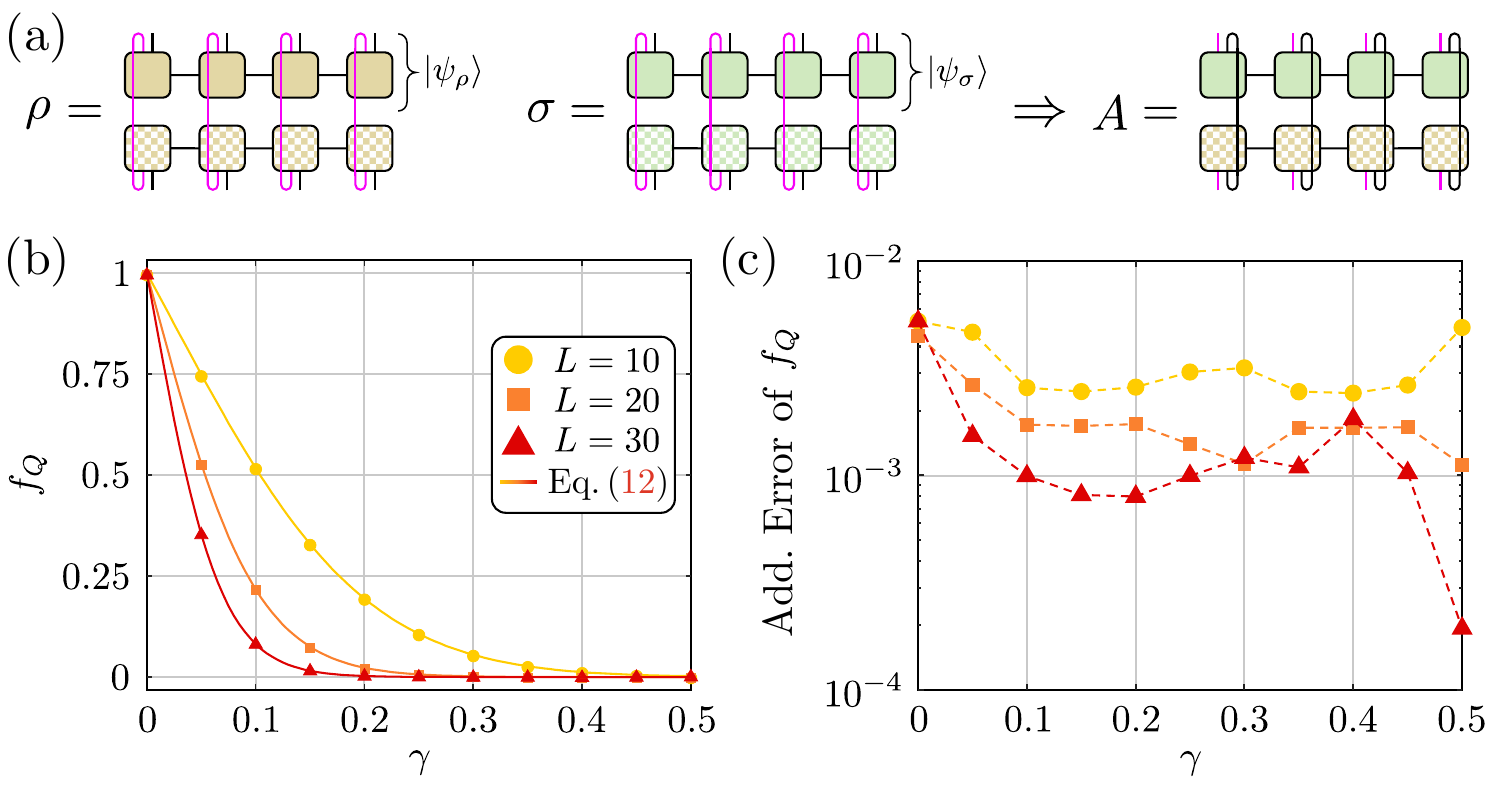}
    \caption{(a) Graphical illustration of the MPO representation of $A = \Tr_{\HH_p} [| \psi_\sigma \rangle \langle \psi_\rho |]$, constructed from the MPS purifications $| \psi_\rho \rangle$ and $| \psi_\sigma \rangle$ of the mixed states $\rho$ and $\sigma$, whose fidelity is to be estimated. Black (magenta) legs of $| \psi_{\rho,\sigma} \rangle$ represent the physical (ancillary) Hilbert space $\HH_p$ ($\HH_a$). (b) Estimated normalized quantum Fisher information $f_Q = F_Q / L^2$ of the GHZ state of length $L$ under amplitude damping noise. The solid curve shows the exact value given in Eq.~\eqref{decoGHZ_qfi}. (c) Additive error of the estimated $f_Q$. The parameters used are $\ell = 10^{-12}$, $r = 30$ and $\chi \leq 20$. For $\theta$, we use $\theta = 0.05$, $0.03$, and $0.025$ for $L = 10$, $20$, and $30$, respectively.}
    \label{fig:fidelity_qfi}
\end{figure}

\textit{Estimating Quantum Fidelity}---Our method extends naturally to estimating the quantum fidelity $F(\rho, \sigma) = \big( \Tr \big[ \sqrt{\rho^{1/2} \sigma \rho^{1/2}} \big] \big)^2$ between two mixed states $\rho$ and $\sigma$, given the MPS representations of their purifications $|\psi_\rho \rangle, |\psi_\sigma \rangle \in \HH_p \otimes \HH_a$ (i.e., $\rho = \Tr_{\HH_a} [|\psi_\rho \rangle \langle \psi_\rho |]$ and $\sigma = \Tr_{\HH_a} [|\psi_\sigma \rangle \langle \psi_\sigma |]$), where $\HH_p$ ($\HH_a$) is the physical (ancillary) Hilbert space. Uhlmann's theorem states that fidelity can be equivalently defined by~\cite{uhlmann1976transition},
\begin{align} \label{UhlmannThm}
    F(\rho, \sigma) = \max_{U \in \mathrm{U}(\HH_a)} | \langle \psi_\rho | (I \otimes U) | \psi_\sigma \rangle |^2,
\end{align}
where $\mathrm{U}(\HH_a)$ is the set of unitary operators on $\HH_a$. Setting $A = \Tr_{\HH_p} [| \psi_\sigma \rangle \langle \psi_\rho |]$, we can rewrite Eq.~\eqref{UhlmannThm} as
\begin{align} \label{Fidelity_rewrite}
    F(\rho, \sigma) = \max_{U \in \mathrm{U}(\HH_a)} | \Tr [A U] |^2 = \| A \|_1^2,
\end{align}
yielding the squared trace norm of $A$. The MPO representation of $A$ is shown graphically in Fig.~\ref{fig:fidelity_qfi}(a). Since $A$ is generally non-Hermitian, we can estimate $\| A \|_1$ by applying our method to the Hermitian matrix $A'$ defined in Eq.~\eqref{NonHerm2Herm}. 

To benchmark our fidelity estimation approach, we apply it to estimate quantum Fisher information (QFI). In quantum metrology, the QFI $F_Q [\rho, H]$ quantifies the ultimate precision limit for estimating a parameter $\theta$ imprinted via unitary encoding $\rho(\theta) := e^{-i\theta H} \rho e^{i\theta H}$, through the quantum Cram\'er-Rao bound~\cite{giovannetti2006quantum,paris2009quantum}. However, direct computation of the QFI for mixed states is generally intractable for large systems. Using the relation $G(\theta) := F(\rho, \rho(\theta)) = 1 - (\theta^2 / 4) F_Q [\rho, H] + \OO(\theta^4)$~\cite{braunstein1994statistical}, the QFI can be approximately extracted from the fidelity between mixed states. To eliminate the leading  bias of order $\OO(\theta^2)$, we evaluate $G(\theta)$ at two values $\theta$ and $2\theta$ via our fidelity estimation method and apply the Richardson extrapolation~\cite{sidi2003practical}, yielding the estimator
\begin{align} \label{qfi_estimator}
    \hat{F}_Q = \frac{4}{3\theta^2} [4(1 - G(\theta)) - (1 - G(2\theta))],
\end{align}
which approximates the $F_Q$ with $\OO(\theta^4)$ error. 

We demonstrate this on a noisy $L$-qubit GHZ state $\rho = \EE_{\mathrm{amp}} (| \psi_{\mathrm{GHZ}} \rangle \langle \psi_{\mathrm{GHZ}} |)$, where $| \psi_{\mathrm{GHZ}} \rangle \propto | 00\dots 0 \rangle + | 11\dots 1 \rangle$ and $\EE_{\mathrm{amp}} = \bigotimes_i \EE_{\mathrm{amp}, i}$ is the amplitude damping channel with error rate $\gamma$ and Kraus operators~\cite{nielsen2001quantum}
\begin{align} \label{amp_damp}
    K_0 = \begin{pmatrix}
        1 & 0 \\
        0 & \sqrt{1 - \gamma}
    \end{pmatrix}, \quad K_1 = \begin{pmatrix}
        0 & \sqrt{\gamma} \\
        0 & 0
    \end{pmatrix}.
\end{align}
The exact QFI of the noisy GHZ state $\rho$ with respect to the observable $H = \frac 12 \sum_i Z_i$ is known analytically~\cite{ma2011quantum}:
\begin{align} \label{decoGHZ_qfi}
    F_Q [\rho, H] = \frac{2L^2 (1 - \gamma)^L}{1 + \gamma^L + (1 - \gamma)^L},
\end{align}
serving as a benchmark. Figures~\ref{fig:fidelity_qfi}(b) and (c) show the estimated normalized QFI $f_Q := F_Q / L^2$ and its relative error, respectively. Our QFI estimation based on Eq.~\eqref{qfi_estimator} matches well the analytic result in Eq.~\eqref{decoGHZ_qfi}, confirming the utility of our method in estimating fidelity.

Several tensor-network approaches have been developed for fidelity estimation. Recently, Ref.~\cite{liu2026polynomial} proposed a variational method that provides lower bounds on the fidelity between mixed states, whereas our approach directly targets the fidelity itself, up to controllable error [see Eq.~\eqref{MyEstimatorErr}]. Earlier methods are limited to pure states~\cite{zhou2008ground} or contiguous subsystems of pure-state MPSs~\cite{hauru2018uhlmann}. In contrast, our method is more broadly applicable: in principle, it can estimate the fidelity between arbitrary mixed states, provided their purifications are available as MPSs, a situation commonly encountered in tensor-network simulations of thermal or open quantum systems. Finally, our method can also be applied to estimate the fidelity correlator, which detects strong-to-weak spontaneous symmetry breaking in mixed states~\cite{lessa2025strong}.  

\textit{Discussion}---In this work, we introduce a tensor-network method for estimating the trace norm of an MPO by leveraging Zolotarev's rational approximation to the sign function. We further show that the same framework extends to estimate the quantum fidelity between two mixed states, provided their purifications are given as MPSs. As benchmarks, we numerically computed the entanglement negativity, fidelity, and quantum Fisher information for various many-body mixed states, which are typically challenging to access with existing tensor-network approaches. 

Combined with recent advances in experimental MPO learning~\cite{votto2026learning}, our method enables the experimental estimation of entanglement negativity for noisy quantum states prepared on noisy quantum processors. Previous experiments were limited to small systems~\cite{liu2023observation} or to measuring moments of the partially transposed density matrix, $\Tr[(\rho^{T_R})^n]$~\cite{elben2020mixed,neven2021symmetry}, which are not genuine entanglement monotones. By combining MPO learning with our approach, direct measurements of entanglement negativity in many-body mixed states may become feasible. It may also provide a tool for benchmarking noise models by computing fidelities from experimentally learned MPOs. Looking ahead, it would be intriguing to extend our method to other tensor-network ansatzes, to higher dimensions, and to other nonlinear quantities amenable to rational approximations. Further improvements in efficiency, e.g., by incorporating symmetries, also constitute a promising direction.

\textit{Acknowledgments---}We thank Changhun Oh, Tarun Grover, Donghoon Kim, Minsoo L. Kim, Wei-Lin Tu, Donggyu Kim and Sarvesh Srinivasan for their helpful discussions. This work was supported by Grant No. KRISS-2026-GP2026-0015 funded by the Korea Research Institute of Standards and Science (KRISS), and by Grant Nos. RS-2024-00451261, RS-2025-00559286, RS-2023-00281839, RS-2023-00256050 and RS-2025-08542968 from the the National Research Foundation of Korea (NRF), funded by the Korean government (Ministry of Science and ICT (MSIT)).

\textit{Data Availability---}The data that support the findings of this article are available from the authors upon reasonable request.

\let\oldaddcontentsline\addcontentsline
\renewcommand{\addcontentsline}[3]{}
\bibliography{ref}
\let\addcontentsline\oldaddcontentsline

\begin{appendix}
\renewcommand{\theequation}{A\arabic{equation}}
\setcounter{equation}{0}
\appendix
\section{End Matter}\label{endmatter}

\textit{Appendix A: Zolotarev's Rational Approximation---}Here we present the explicit formula for Zolotarev's rational approximation $Z_{\ell,r} (x)$ to the sign function on $[-1, -\ell] \cup [\ell, 1]$. For convenience, we repeat its definition~\cite{zolotarev1877application,achieser2013theory,nakatsukasa2016computing}:
\begin{align} \label{Zolotarev2}
    Z_{\ell,r} (x) := M x \left( 1 + \sum_{j=1}^r \frac{a_j}{x^2 + c_{2j-1}} \right).
\end{align}
The coefficients $c_i$ are given by
\begin{align} \label{c}
    c_i = \ell^2 \frac{\mathrm{sn} \left( \frac{i K'}{2r + 1}; \ell' \right)}{\mathrm{cn} \left( \frac{i K'}{2r + 1}; \ell' \right)}, \quad 1\leq i\leq 2r,
\end{align}
where $\ell' = \sqrt{1 - \ell^2}$, $K' = \int_0^{\pi/2} \frac{d\theta}{\sqrt{1 - \ell'^2 \sin^2 \theta}}$, and $\mathrm{sn}(u; \ell')$ and $\mathrm{cn}(u; \ell')$ denote the Jacobi elliptic functions. The coefficients $a_j$ are 
\begin{align} \label{a}
    a_j = \frac{\prod_{k = 1}^r (c_{2k} - c_{2j-1})}{\prod_{k = 1, k\neq j}^r (c_{2k - 1} - c_{2j-1})}, \quad 1\leq j\leq r,
\end{align}
and the normalization constant $M$ is
\begin{align} \label{M}
    M = \frac{2}{\prod_{j=1}^r \frac{1 + c_{2j}}{1 + c_{2j - 1}} + \ell \prod_{j=1}^r \frac{\ell^2 + c_{2j}}{\ell^2 + c_{2j - 1}}}.
\end{align}

The rational function $Z_{\ell,r} (x)$ is the unique solution of the min-max problem
\begin{align} \label{minmax}
    \min_{R \in \mathcal{R}_{2r+1, 2r}} \max_{x\in [-1,-\ell] \cup [\ell, 1]} | \sgn(x) - R(x) |,
\end{align}
where $\mathcal{R}_{2r+1, 2r}$ denotes the set of all rational functions $R = P / Q$, with $P$ and $Q$ real polynomials of degree at most $2r+1$ and $2r$, respectively. The approximation error is known to decay exponentially with $r$:
\begin{align} \label{ZolotarevErr2}
    \max_{x\in [-1,-\ell] \cup [\ell, 1]} | \sgn(x) - Z_{\ell,r} (x) | = C_{r} \rho^{-(2r+1)}.
\end{align}
where $\rho = e^{\pi K(\mu') / 4 K(\mu)} > 1$, $\mu = \frac{1 - \sqrt{\ell}}{1 + \sqrt{\ell}}$, $\mu' = \sqrt{1 - \mu^2}$, $K(x) = \int_0^{\pi/2} \frac{d\theta}{1 - x^2 \sin^2 \theta}$, and $1 \leq \frac{2}{1 + \rho^{-(2r+1)}} \leq C_r \leq \frac{2}{1 - \rho^{-(2r+1)}} \leq \frac{2}{1 - \rho^{-1}}$.

\vspace{10pt}
\textit{Appendix B: Derivation of the Relative Error Bound---}Here we derive the error bound Eq.~\eqref{MyEstimatorErr}. We assume that $A$ is Hermitian and that $|A| \preceq 1$. The absolute error can be written as
\begin{align}
    \left| f_{\ell,r} (A) - \| A \|_1 \right| &= \left| \Tr[A \cdot (Z_{\ell,r}(A) - \sgn(A))]\right| \\
    &\leq \sum_i |\lambda_i| \cdot \left| \sgn(\lambda_i) - Z_{\ell,r} (\lambda_i) \right|, \nonumber
\end{align}
where $\lambda_i$ are the eigenvalues of $A$. Using Eq.~\eqref{ZolotarevErr2} for $\ell \leq |\lambda_i| \leq 1$ and $| \sgn(\lambda_i) - Z_{\ell,r} (\lambda_i) | \leq 1$ for $|\lambda_i| < \ell$, we obtain
\begin{align}
    \left| f_{\ell,r} (A) - \| A \|_1 \right| &\leq \sum_{i:|\lambda_i| < \ell} |\lambda_i| + \sum_{i:\ell \leq |\lambda_i| \leq 1} |\lambda_i| C_r \rho^{-(2r+1)} \nonumber \\
    &\leq \sum_{i:|\lambda_i| < \ell} |\lambda_i| + \| A \|_1 C_r \rho^{-(2r+1)} \\
    &\leq \|A\|_1 \left( C_r \rho^{-(2r+1)} + R_{<\ell} \right), \nonumber
\end{align}
where $R_{<\ell} \equiv \|A\|_1^{-1} \sum_{i:|\lambda_i| < \ell} |\lambda_i|$. Therefore, the relative error bound in Eq.~\eqref{MyEstimatorErr} directly follows.

\end{appendix}

\clearpage
\onecolumngrid

\renewcommand{\thefigure}{S\arabic{figure}}
\renewcommand{\theequation}{S\arabic{equation}}
\renewcommand{\thetable}{S\Roman{table}}
\renewcommand{\thesection}{S\Roman{section}}
\renewcommand{\thesubsection}{\Alph{subsection}}
\renewcommand{\thesubsubsection}{\arabic{subsubsection}}

\newpage

\centerline{\large{\textbf{Supplementary Material: Estimating the Trace Norm of Matrix Product Operators}}}
\vspace{10pt}

\centerline{Seunghun Lee and Eun-Gook Moon}
\centerline{\textit{Department of Physics, Korea Advanced Institute of Science and Technology, Daejeon 34141, Republic of Korea}}

\setcounter{equation}{0}
\setcounter{figure}{0}
\setcounter{secnumdepth}{0}
\setcounter{theorem}{0}
\setcounter{lemma}{0}

\vspace{0.5cm}

\section{Entanglement Negativity of Decohered Cluster State}

In this section, we derive Eq.~\eqref{Neg_DecoCluster} in the main text, which gives an analytic expression for the half-chain entanglement negativity of the 1D cluster state under $Z$ noise. Specifically, the cluster state $| \psi \rangle$ on an open chain is the ground state of the Hamiltonian $H = -\sum_{i=1}^L K_i$, where $K_1 = X_1 Z_2$, $K_L = Z_{L-1} X_L$, and $K_i = Z_{i-1} X_i Z_{i+1}$ for $2 \leq j \leq L - 1$~\cite{briegel2001persistent}. For simplicity, we assume that $L$ is even. We consider the mixed state $\rho = \EE_Z [| \psi \rangle \langle \psi |]$ which results from the action of the local phase-flip channel $\EE_Z = \bigotimes_i \EE_{Z, i}$, where $\EE_{Z, i}[\cdot] = (1-p) (\cdot) + p Z_i (\cdot) Z_i$ and $p \in [0, 1/2]$ is an error rate. 

The derivation proceeds along the lines of Ref.~\cite{lee2025robust}. Since the cluster state is a stabilizer state obeying $K_j | \psi \rangle = | \psi \rangle$, its density matrix can be expressed as $| \psi \rangle \langle \psi | \propto \sum_{\{a\}} \prod_i K_i^{a_i}$, where $a_i \in \{0, 1\}$ indicates the presence or absence of the stabilizer $K_i$. Under conjugation by $Z_k$, each term $\prod_i K_i^{a_i}$ flips (maintains) its sign when $a_i = 1$ ($0$). As a result, the spectrum of the decohered density matrix $\rho$ can be written as
\begin{align}
    \rho (\{ K \}) \propto \sum_{\{a\}} \prod_i K_i^{a_i} (1 - 2p)^{\sum_i a_i}.
\end{align}
Let $R$ denote the half-space subsystem. To take the partial transpose with respect to $R$, we need to compute $\left[ \prod_i K_i^{a_i} \right]^{T_R}$. Since $Y^T = -Y$ ($Y$: Pauli-$Y$ operator), the partial transpose on $R$ introduces a minus sign only when $a_{L/2} = a_{L/2+1} = 1$. Consequently, we obtain $\rho^{T_R} (\{ K \}) = C^{-1} \Lambda_1 (\{ K \}) \Lambda_2 (\{ K \})$, where
\begin{equation}
    \Lambda_1 (\{ K \}) = \prod_{i\neq L/2, L/2+1} \sum_{a_i = 0}^1 K_i^{a_i} (1 - 2p)^{a_i} = \prod_{i\neq L/2, L/2+1} \left[ 1 + K_i (1 - 2p) \right] \geq 0,
\end{equation}
and
    \begin{equation}
        \begin{aligned}
            \Lambda_2 (\{ K \}) &= \sum_{a_{L/2}, a_{L/2 + 1} = 0}^1 K_{L/2}^{a_{L/2}} K_{L/2 + 1}^{a_{L/2 + 1}} (1 - 2p)^{a_{L/2} + a_{L/2 + 1}} (-1)^{a_{L/2} a_{L/2 + 1}} \\
            &= \begin{cases}
                2 - 4p^2 & \text{if } K_{L/2} = K_{L/2 + 1} = 1, \\
                2 - 4p + 4p^2 & \text{if } K_{L/2} = 1,\, K_{L/2 + 1} = -1, \\
                2 - 4p + 4p^2 & \text{if } K_{L/2} = -1,\, K_{L/2 + 1} = 1, \\
                -2 + 8p - 4p^2 & \text{if } K_{L/2} = K_{L/2 + 1} = -1.
            \end{cases}
        \end{aligned}
    \end{equation}
The normalization constant is $C = 2^L$. Now, observe that $\Lambda_2 < 0$ only when $K_{L/2} = K_{L/2 + 1} = -1$ and $0 \leq p < p_c$, where $p_c = 1 - 1/\sqrt{2}$. Therefore, for $p_c \leq p$, $\rho^{T_R}$ has no negative eigenvalues, and the entanglement negativity vanishes. On the other hand, for $0 \leq p < p_c$, the negativity $\NN = \big|\sum_{\{K\}: \rho^{T_R} (\{K\}) < 0} \rho^{T_R} (\{K\}) \big|$ evaluates to
\begin{align}
    \NN(p) = \frac{4p^2 - 8p + 2}{C} \prod_{j\neq L/2, L/2+1} \sum_{K_j = \pm 1} \Lambda_1 (\{ K \}) = p^2 - 2p + \frac 12.
\end{align}
In summary, the entanglement negativity is given by
\begin{align}
    \NN(p) = \left( p^2 - 2p + \frac 12 \right) \Theta (p_c - p),
\end{align}
which reproduces Eq.~\eqref{Neg_DecoCluster} in the main text. Here, $\Theta(\cdot)$ denotes the Heaviside theta function.

\section{Convergence with Bond Dimension}

\begin{figure*}[b]
    \centering
    \includegraphics[width=\columnwidth]{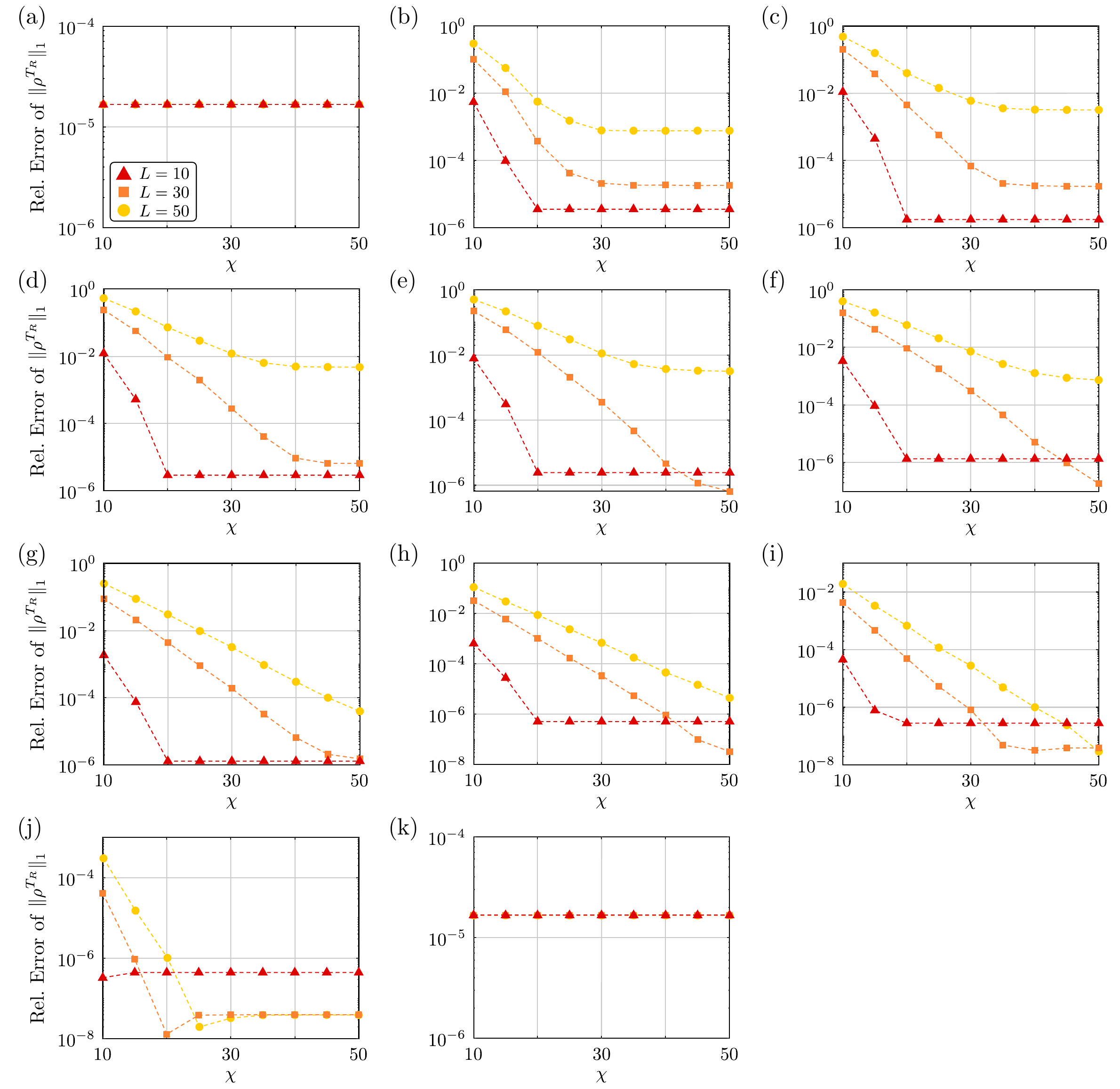}
    \caption{Relative error of the estimated trace norm $\| \rho^{T_R} \|_1$ for the $Z$-decohered cluster state (with $R$ denoting the half-chain subsystem), shown as a function of the bond dimension $\chi$ of the variational MPS $|X\rangle$. Panels (a)–(k) correspond to (a) $p = 0$, (b) $p = 0.05$, (c) $p = 0.1$, (d) $p = 0.15$, (e) $p = 0.2$, (f) $p = 0.25$, (g) $p = 0.3$, (h) $p = 0.35$, (i) $p = 0.4$, (j) $p = 0.45$, and (k) $p = 0.5$, respectively. The parameters used are $\ell = 10^{-10}$ and $r = 30$.}
    \label{fig:DecoCluster_Neg_Conv_chi}
\end{figure*}

In this section, we examine the convergence of our method with respect to the bond dimension $\chi$ of the variational MPS $| X \rangle$. Since the required $\chi$ is not known \textit{a priori}, we increase $\chi$ sequentially, using the converged MPS at each $\chi$ as the initial MPS for the next. The relative error as a function of $\chi$ is shown in Fig.~\ref{fig:DecoCluster_Neg_Conv_chi} for the partially transposed density matrix of the $Z$-decohered cluster state. The error decreases and eventually saturates as $\chi$ increases, demonstrating the controlled convergence of our approach. The plataeus at large system size $L$ is attributed to the nonzero $R_{< \ell}$ contribution from small eigenvalues within $(-\ell, \ell)$, which can be reduced by choosing a smaller $\ell$. 

\section{Comparison with MPO Lanczos Method}

In this section, we compare our trace-norm estimation method with the global Lanczos algorithm of Refs.~\cite{august2017on,august2018efficient}, which approximates $\Tr[f(A)]$ of an MPO $A$ for a smooth function $f$. While one may formally apply it to $f(x) = |x|$ to estimate $\| A \|_1 = \Tr [|A|]$, the lack of smoothness of $f(x) = |x|$ at $x = 0$ leads to several limitations.

The best polynomial approximation to $|x|$ on $[-1, 1]$ converges only algebraically, i.e.,
\begin{align}
    \min_{p \in P_K} \max_{x\in [-1, 1]} ||x| - p(x)| = \Theta(1/K),
\end{align}
where $P_K$ denotes the set of all polynomials $p$ with $\deg(p) \leq K$ (see Theorem 7.2.1 of Ref.~\cite{timan2014theory}). In contrast, Zolotarev's rational approximation $Z_{\ell,r}$ achieves exponential convergence in $r$ by restricting to the domain $[-1, -\ell] \cup [\ell, 1]$ (see Eq.~\eqref{MyEstimatorErr} of the main text). This has the advantage of significantly reducing the number of terms required to achieve a target accuracy $\epsilon$ from $\OO(1/\epsilon)$ in the polynomial case to $\OO(\log(1/\epsilon))$ for Zolotarev's approximation. Moreover, the Zolotarev approach provides an explicit and rigorous error bound that is controllable by tuning the parameters $r$ and $\ell$, whereas the Lanczos method offers no such guarantee for non-smooth functions like $f(x) = |x|$. Finally, since the Lanczos method is based on polynomial approximation of $f$, its slow algebraic convergence necessitates a large Kryloc dimension $K$. In the MPO setting, this can lead to a significant accumulation of truncation errors from bond-dimension growth and compression in MPO--MPO additions and multiplications.

\begin{figure}[h]
    \includegraphics[width=0.77\columnwidth]{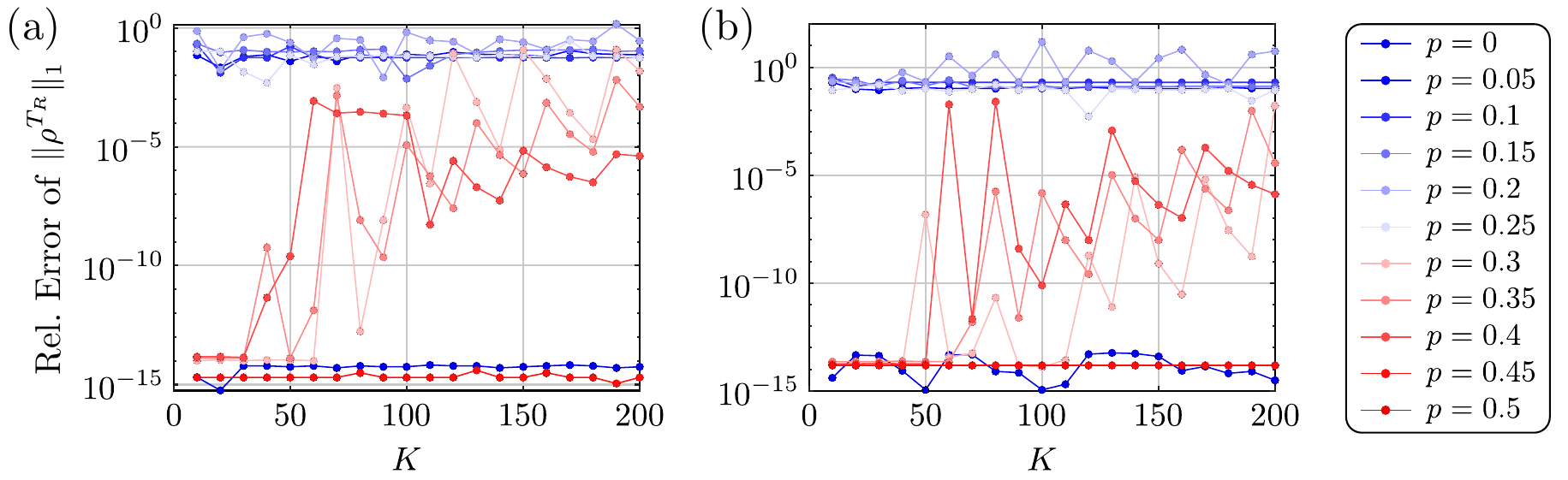}
    \caption{Relative error of the estimated trace norm $\| \rho^{T_R} \|_1$ for the $Z$-decohered cluster state using the MPO global Lanczos algorithm of Refs.~\cite{august2017on,august2018efficient}, shown as a function of the Krylov dimension $K$. Panels (a) and (b) correspond to system sizes $L = 30$ and $L = 50$, respectively. The maximum bond dimension used is $D_{\mathrm{max}} = 100$.}
    \label{fig:DecoCluster_Neg_Lanczos_Comparison}
\end{figure}

We benchmark the global Lanczos algorithm on computing entanglement negativity of $Z$-decohered cluster state, with results shown in Fig.~\ref{fig:DecoCluster_Neg_Lanczos_Comparison}. We find that the relative error does not decrease as the Krylov dimension $K$ increases when the density matrix is not a stabilizer state and its partial transpose has a significant number of negative eigenvalues, i.e., when $0 < p < p_c \simeq 0.293$. Furthermore, truncation errors grow with increasing $K$ for some error rates. This is in contrast with our Zolotarev method, consistently achieves higher and controllable accuracy in trace-norm estimation. 

\end{document}